\documentclass[aip,jap,preprint,tightenlines,showpacs]{revtex4-1}
\usepackage{graphicx}% Include figure files
\usepackage{bm}% bold math
\usepackage{amsmath}
\usepackage{amssymb}
\usepackage{dcolumn}
\newcommand{\APL}{Appl. Phys. Lett. }
\newcommand{\APNY}{Ann. Phys. (N.Y.) }
\newcommand{\JAP}{J. Appl. Phys. }
\newcommand{\JLum}{J. Lumin. }
\newcommand{\PhysE}{Physica E (Amsterdam) }
\newcommand{\SM}{Superlatt. Microstruct. }
\newcommand{\PR}{Phys. Rev. }

\newcommand{\PRB}{Phys. Rev. B }

\newcommand{\PRL}{Phys. Rev. Lett. }
\newcommand{\SST}{Semicond. Sci. Technol. }
\newcommand{\IJMPB}{Int. J. Mod. Phys. B }
\newcommand{\EPL}{Europhys. Lett. }
\newcommand{\SSC}{Solid State Commun. }
\newcommand{\JPCM}{J. Phys.: Condens. Matter }
\newcommand{\OC}{Opt. Commun. }

\begin{document}
\title{Magnetic field control of the intraband optical absorption in two-dimensional quantum rings}
\author{O.~Olendski}
\email{oolendski@ksu.edu.sa}
\affiliation{King Abdullah Institute for Nanotechnology, King Saud University, P.O. Box 2454, Riyadh 11451, Saudi Arabia}
\author{T.~Barakat}
\email{tbarakat@ksu.edu.sa}
\affiliation{Department of Physics, King Saud University, P.O. Box 2454, Riyadh 11451, Saudi Arabia}
\date{\today}
\begin{abstract}
Linear and nonlinear optical absorption coefficients of the two-dimensional semiconductor ring in the perpendicular magnetic field $\bf B$ are calculated within independent electron approximation. Characteristic feature of the energy spectrum are crossings of the levels with adjacent nonpositive magnetic quantum numbers as the intensity $B$ changes. It is shown that the absorption coefficient  of the associated optical transition is drastically decreased at the fields corresponding to the crossing. Proposed model of the Volcano disc allows to get simple mathematical analytical results, which provide clear physical interpretation. An interplay between positive linear and intensity-dependent negative cubic absorption coefficients is discussed; in particular, critical light intensity at which additional resonances appear in the total absorption dependence on the light frequency is calculated as a function of the magnetic field and levels' broadening.
\end{abstract}
\vskip.7in
\maketitle

\section{Introduction}\label{sec_1}
Nonsimply connected topology of the quantum rings has been attracting for a long time a careful attention of physicists, chemists, mathematicians. Experimentally, impressive successes of the growth technologies in the past years allowed to build up structures of almost any desired shape. As a result, the Aharonov-Bohm \cite{Aharonov1} (AB) oscillations were observed in the electronic semiconductor quantum rings \cite{Liu2,Fuhrer1,Keyser1,Giesbers1,Kleemans1} and structures with the quantum antidot \cite{Sachrajda1,Goldman1} (QAD). They were also detected for the hole rings.\cite{Yau1,Habib1} Recently, complex quantum ring structures were fabricated by the droplet epitaxy with varying temperature \cite{Huang1} or local droplet etching.\cite{Stemmann1} On the other hand, a correct  understanding of the properties of these ultra-small man-made nanostructures requires an adequate choice of the theoretical models. Features of the combination of the coaxial parabolic and inverse parabolic potentials, $U(\rho)\sim a_1\rho^2+a_2/\rho^2$, that is called a Volcano disk (VD) have been amply calculated in recent investigations, \cite{Bogachek1,Tan1,Tan2,Tan3,Fukuyama1,Simonin1,Bulaev1,Margulis1} especially when it is placed in the uniform magnetic field $\bf B$. Such a simple geometry for nonzero $a_i$, $i=1,2$, describes an isolated ring of the finite width and average radius $\rho_V=\sqrt[4]{a_2/a_1}$. \cite{Tan1,Tan2,Tan3,Bulaev1} Beside quantum ring, a flexibility of the model allows, by the variation of  the parameters $a_i$, \cite{Tan2} to describe a quantum dot \cite{Tanaka1} (QD) if $a_2=0$, an isolated QAD \cite{Bogachek1} ($a_1=0$), one-dimensional ring [when $\rho_V={\rm const}$ and the value of $\omega_V=\sqrt{8a_1/m^*}$ ($m^*$ is carrier effective mass) tends to infinity, $\omega_V\rightarrow\infty$] or two-dimensional straight wire ($\omega_V={\rm const}$ and $\rho_V\rightarrow\infty$). Among other models of quantum rings and QDs, \cite{Simonin1,Halonen1,Chakraborty1,Hu1,Hu2,Govorov1,Fomin1,Teodoro1,Arsoski1,Voskoboynikov1} the most popular is the displaced parabola potential of the form
\begin{equation}\label{DisplacedParabola1}
U(\rho)=\frac{1}{2}m^\ast\omega_0^2(\rho-\rho_D)^2
\end{equation}
with the nonzero displacement $\rho_D$. A comparison of these two models shows that the divergence of the VD  potential at $\rho=0$ forbids the particle presence in the origin what means that the structure possesses strictly doubly connected geometry. Similar topology is produced also by the flat potential with the infinite inner and outer hard-wall confinement;\cite{Avishai1} however, contrary to these models, the VD representation allows exact simple analytical solutions for the energy spectra and wave functions. \cite{Tan1,Tan2,Tan3,Bulaev1} It was shown \cite{Tan1,Tan3} that such a model, despite of its simplicity, correctly explains experimental data, in particular, the beating effect in the  oscillation pattern \cite{Liu2} and a magnitude of a persistent current \cite{Buttiker1} in a GaAs/Al$_x$Ga$_{1-x}$As single loop. \cite{Mailly1}

Previous analysis concentrated mainly on the calculation of the transport and thermodynamic properties of the ring \cite{Bogachek1,Tan1,Tan2,Tan3,Fukuyama1,Simonin1,Bulaev1,Margulis1} while the results on the optical response are scarce \cite{Halonen1,Climente2} what can be explained (at least, partially) by the huge technological challenges of the far-infrared (FIR) measurements. Experimental overcoming of these difficulties \cite{Lorke1,Warburton1} stimulates further theoretical research on the optics of the rings.

Below, in the framework of the effective-mass approximation and independent electron formalism we provide analysis of the linear and nonlinear absorption coefficients of the quantum ring with its potential modelled by the Volcano form. Note that we study electron optical transitions inside the conduction band (intraband transitions) what makes the difference as compared to the AB effect for excitons \cite{Hu2,Govorov1,Teodoro1,Arsoski1} where the optical recombination between  the electron in the conduction band and hole in the valence band (interband process) plays an essential role. The model of the VD allows to get clear  analytical equations, which reveal, among other results, the magnitude of the field at which the two levels with adjacent magnetic quantum numbers cross and its dependence on the parameters of the ring; in particular, for the very small width of the loop the distance between the crossings on the $B$ axis is determined by the change of the flux through it by the one flux quantum. It is shown that the linear absorption drastically decreases with the intensity $\bf B$ approaching this value and the physical reason for this is given. The interplay between the linear and cubic contributions to the total absorption is discussed and the critical optical intensity, which switches emergence of the new resonances in the optical spectrum, is calculated and analysed. Generalization of this result to any system where the influence of the nonlinear term becomes noticeable is given.

The outline of the paper is as follows. A formulation of the problem is presented in Sec.~\ref{sec_2}.  Sec. ~\ref{sec_3} discusses the obtained results. Some concluding remarks are given in Sec.~\ref{sec_4}.

\section{Model and Formulation}\label{sec_2}
We consider a quantum ring grown on the flat surface with its lateral dimension in the direction perpendicular to the interface being much smaller than the outer radius of the annulus.  Then, the motion of the particle becomes essentially two-dimensional (2D). The form of the function describing the  variation of the ring potential is chosen to be the one of the VD, which in the polar coordinates ${\bm\rho}\equiv(\rho,\varphi)$ is given by 
\begin{equation}\label{Potential1}
U(\rho,\varphi)\equiv U(\rho)=\frac{1}{2}m^*\omega_0^2\rho^2+\frac{\hbar^2}{2m^*\rho^2}\,a-\hbar\omega_0a^{1/2}.
\end{equation}
Here, $m^*$ is, as mentioned above, an effective mass of a charge carrier, frequency $\omega_0$ defines a steepness of the confining in-plane surface of the QD with its effective radius $\rho_0=[\hbar/(2m^*\omega_0)]^{1/2}$, and positive dimensionless constant $a$ describes a strength of the repulsive potential of the QAD while the constant last term in the right-hand-side is introduced for the compensation purposes since it reduces the minimum of the potential to the zero value. This sole extremum of $U(\rho)$ is achieved at 
\begin{equation}\label{Radius1}
\rho_V=2^{1/2}a^{1/4}\rho_0,
\end{equation}
which can be considered as a mean radius of the Volcano ring. In turn, energy $E$-dependent innner $\rho_{in}$ and outer $\rho_{out}$ radii are given as
%
%\begin{subequations}\label{InnerOuterRadii}
%\begin{eqnarray}\label{InnerRarius1}
%\rho_{in}&=&2^{1/2}\left[a^{1/2}+\frac{E}{\hbar\omega_0}-\sqrt{\left(a^{1/2}+\frac{E}{\hbar\omega_0}\right)^2-a}\right]^{1/2}\rho_0\\
%\label{OuterRadius1}
%\rho_{out}&=&2^{1/2}\left[a^{1/2}+\frac{E}{\hbar\omega_0}+\sqrt{\left(a^{1/2}+\frac{E}{\hbar\omega_0}\right)^2-a}\right]^{1/2}\rho_0.
%\end{eqnarray}
%\end{subequations}
%
%
\begin{subequations}\label{InnerOuterRadii}
\begin{eqnarray}\label{InnerRarius1}
\rho_{in}&=&2^{1/2}\negthickspace\left(a^{1/2}+\frac{E}{\hbar\omega_0}-\left[\left(a^{1/2}+\frac{E}{\hbar\omega_0}\right)^2-a\right]^{1/2}\right)^{\negthickspace 1/2}\negthickspace\rho_0\\
\label{OuterRadius1}
\rho_{out}&=&2^{1/2}\negthickspace\left(a^{1/2}+\frac{E}{\hbar\omega_0}+\left[\left(a^{1/2}+\frac{E}{\hbar\omega_0}\right)^2-a\right]^{1/2}\right)^{\negthickspace 1/2}\negthickspace\rho_0.
\end{eqnarray}
\end{subequations}

\begin{figure}
\centering
\includegraphics[width=0.8\columnwidth]{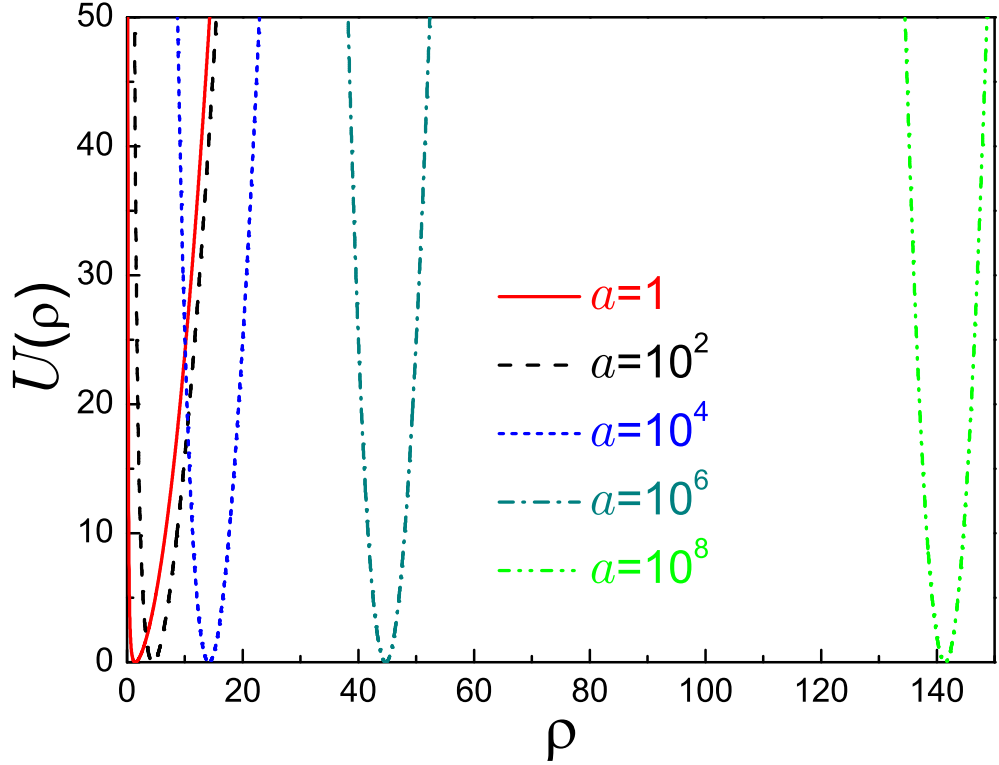}
\caption{\label{Fig1}
Potential profile $U(\rho)$ (in units of $\hbar\omega_0$) as a function of the radius $\rho$ (in units of $\rho_0$) for $a=1$ (solid line), $a=10^2$ (dashed curve), $a=10^4$ (dotted curve), $a=10^6$ (dash-dotted line) and $a=10^8$ (dash-dot-dotted curve).}
\end{figure}

Asymptotic cases of these equations are
\begin{subequations}\label{InnerOuterRadii2}
\begin{eqnarray}\label{InnerRarius2}
\rho_{in}&=&\rho_0\left\{\negthickspace\negthickspace
\begin{array}{cc}
\left(\,a\hbar\omega_0/E\right)^{1/2}, &a\ll 1\\
\,2^{1/2}a^{1/4}+\!E/\!\left(2^{1/2}a^{1/4}\hbar\omega_0\right)-\!\left[E/\!\left(\hbar\omega_0\right)\right]^{1/2},&a\gg 1,
\end{array}
\right.\\
\label{OuterRadius2}
\rho_{out}&=&\rho_0\left\{\negthickspace\negthickspace
\begin{array}{cc}
2\left[a^{1/2}+E/\!\left(\hbar\omega_0\right)\right]^{1/2}, &a\ll 1\\
\,2^{1/2}a^{1/4}+\!E/\!\left(2^{1/2}a^{1/4}\hbar\omega_0\right)+\!\left[E/\!\left(\hbar\omega_0\right)\right]^{1/2},&a\gg 1.
\end{array}
\right.
\end{eqnarray}
\end{subequations}
It is seen from Eqs.~\eqref{Radius1}--\eqref{InnerOuterRadii2} that in our case the natural and the most convenient units of measuring energy, frequency and distance are $\hbar\omega_0$, $\omega_0$ and $\rho_0$, respectively. We will frequently use this fact below  during discussion of the obtained results. They also show that by varying parameters $a$ and $\omega_0$ and the Fermi Energy $E$ one can model rings of the different shapes. Profile of the potential $U(\rho)$ is shown in Fig.~\ref{Fig1} for several antidot strengths $a$. It is seen that the VD with the small $a$ can be considered as a ``thick" ring while the one with the large antidot strength generally describes quite well a ``thin" annulus. \cite{Tan3}
In addition, a uniform magnetic field ${\bf B}=(0,0,B)$ is applied perpendicularly to the 2D plane.  Then, in the Schr\"{o}dinger equation 
\begin{equation}\label{Schrodinger1}
\hat{H}\Psi(\bm\rho)=E\Psi(\bm\rho)
\end{equation} 
for finding eigenenergies $E$ and corresponding eigenfunctions $\Psi(\bm\rho)$, the Hamiltonian $\hat{H}$ is written as
\begin{equation}\label{Hamiltonian1}
\hat{H}=\frac{1}{2m^*}(-i\hbar{\bm\nabla}+e{\bf A})^2+U(\bm\rho)
\end{equation}
with $e$ being an absolute value of the electronic charge. Magnetic field enters the equation through the vector potential $\bf A$: ${\bf B}={\bm\nabla}\times{\bf A}$. Here, it is convenient to choose a symmetric gauge where the 2D vector potential $\bf A$ takes the form ${\bf A}=(0,B\rho/2)$. Then, the energies $E$ form a countably infinite set \cite{Bogachek1,Tan1,Tan2,Tan3,Bogachek1,Olendski1}
\begin{equation}\label{Energy0}
E_{a;nm}\equiv E_{a;nm}(\omega_c)=\hbar\omega_{e\!f\!f}\left(2n+\sqrt{m^2+a}+1\right)+\frac{1}{2}m\hbar\omega_c-\hbar\omega_0a^{1/2}.
\end{equation}
Here, $n=0,1,2,\ldots$ and $m=0,\pm 1,\pm 2,\ldots$ are the principal and azimuthal quantum numbers, respectively;  $\omega_c=eB/m^*$ is the cyclotron frequency, and $\omega_{e\!f\!f}=(\omega_0^2+\omega_c^2/4)^{1/2}$. Wave functions $\left|n,m\right>$ corresponding to the energies, Eq.~(\ref{Energy0}), are 
\begin{equation}\label{Wavefunction0}
\left|n,m\right>\equiv\Psi_{nm}(a;\rho,\varphi)=\frac{e^{im\varphi}}{(2\pi)^{1/2}}R_{nm}(a;\rho)
\end{equation}
with the radial dependencies $R_{nm}(a;\rho)$ expressed as 
\begin{eqnarray}
R_{nm}(a;\rho)=\frac{1}{\rho_{e\!f\!f}}\left[\frac{n!}{\Gamma(n+\sqrt{m^2+a}+1)}\right]^{1/2}\nonumber\\
\label{Radial0}
\times\exp\left(-\frac{1}{4}\frac{\rho^2}{\rho_{e\!f\!f}^2}\right)\left(\frac{1}{2}\frac{\rho^2}{\rho_{e\!f\!f}^2}\right)^{\sqrt{m^2+a}/2}L_n^{\sqrt{m^2+a}}\left(\frac{1}{2}\frac{\rho^2}{\rho_{e\!f\!f}^2}\right).
\end{eqnarray}
Here, $\rho_{e\!f\!f}=[\hbar/(2m^*\omega_{e\!f\!f})]^{1/2}$, $\Gamma(x)$ is $\Gamma$-function and $L_n^\alpha(x)$ is an associated Laguerre polynomial. \cite{Abramowitz1} Equation~\eqref{Radial0} shows that for the ring structure, $a\neq0$, the electron can not be found in the origin for any quantum numbers $n$ and $m$, as it was already mentioned in the Introduction: $\Psi_{nm}(a\neq0;\rho=0,\varphi)=0$. Functions $\Psi_{nm}(a;\rho,\varphi)$ are orthonormalized according to
\begin{equation}\label{normalization0}
\int d{\bm\rho}\Psi_{n'm'}^*(a;\rho,\varphi)\Psi_{nm}(a;\rho,\varphi)=\delta_{nn'}\delta_{mm'},
\end{equation}
$\delta_{nn'}$ is a Kronnecker symbol. Obviously, equations above for $a=0$ simplify to their QD counterparts, \cite{Tanaka1} where the circularly symmetric states, $m=0$, have the largest electron concentration at $\rho=0$.

\begin{figure}
\centering
\includegraphics[width=0.8\columnwidth]{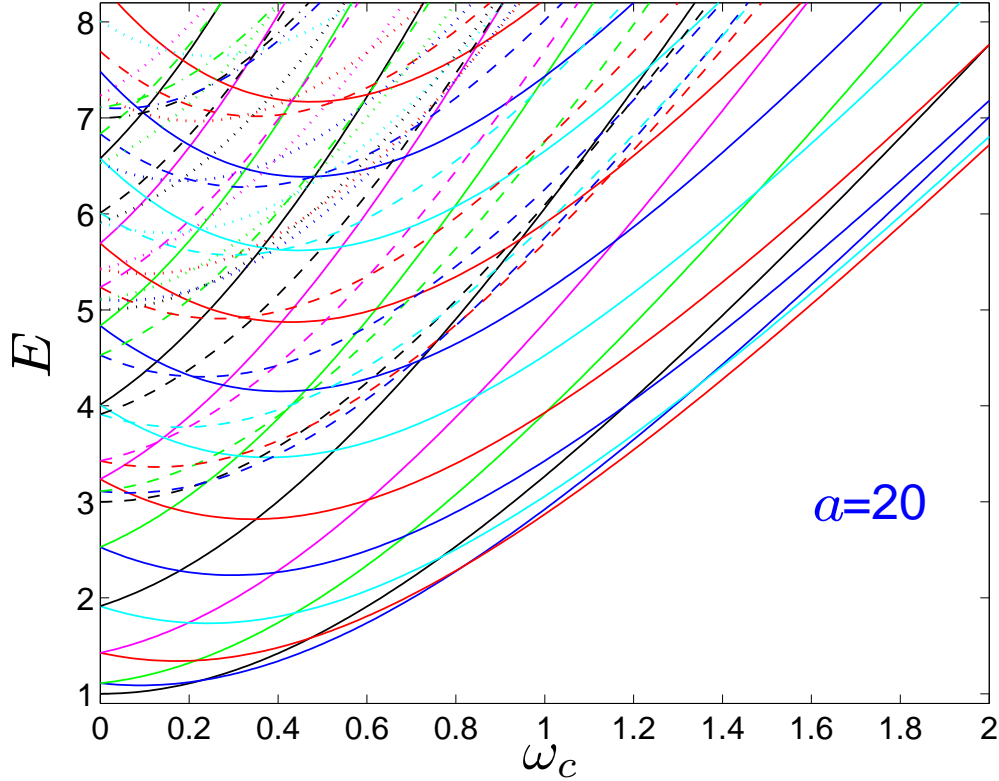}
\caption{\label{Fig2}
Energies $E_{a;nm}$ (in units of $\hbar\omega_0$) as a function of the cyclotron frequency $\omega_c$ (in units of $\omega_0$) of the VD with $a=20$. Solid lines show a family of levels with $n=0$ and different $m$, dashed lines are the states with $n=1$, dotted curves are the levels with $n=2$, and dash-dotted lines - $n=3$. Levels $\left|n,0\right>$ and $\left|n,-1\right>$ cross at $\omega_0^\times=0.2222$.}
\end{figure}

Energy spectrum is shown in Fig. \ref{Fig2} as a function of the magnetic field for the strength $a=20$. It shows that in the absence of the magnetic fields the energies of the radially symmetric states are equal to the odd integers of the fundamental confining energy
\begin{equation}\label{EnergyAsymptotics1}
E_{a;n0}(\omega_c=0)=(2n+1)\hbar\omega_0,
\end{equation}
as it directly follows from Eq.~\eqref{Energy0}. The most prominent feature in Fig.~\ref{Fig2} are crossings of the levels with the same $n$ and different nonpositive $m$ as $\omega_c$ growing. For example, levels $\left|n,-|m|\right>$ and $\left|n,-|m|-1\right>$ cross at $\omega_m^\times$ equal to
\begin{equation}\label{Crossing0}
%\omega_m^\times=\frac{2\left(\sqrt{\left(|m|+1\right)^2+a}-\sqrt{|m|^2+a}\right)}{\left[1-\left(\sqrt{\left(|m|+1\right)^2+a}-\sqrt{|m|^2+a}\right)^2\right]^{1/2}}\,\omega_0,
\omega_m^\times=2\frac{\sqrt{\left(|m|+1\right)^2+a}-\sqrt{|m|^2+a}}{\left[1-\left(\sqrt{\left(|m|+1\right)^2+a}-\sqrt{|m|^2+a}\right)^2\right]^{1/2}}\,\omega_0,
\end{equation}
which is independent of the radial quantum number $n$. Asymptotic limits of Eq.~\eqref{Crossing0} are
\begin{equation}\label{Crossing1}
\omega_m^\times=\omega_0\left\{\negthickspace\negthickspace\negthickspace
\begin{array}{cc}
\left.\begin{array}{cc}
2^{1/2}/a^{1/4},&m=0\\
2\left[\,|m|\left(|m|+1\right)/a\right]^{1/2},&m<0
\end{array}
\right\}&a\ll 1\\
(2|m|+1)/a^{1/2},&a\gg 1.
\end{array}
\right.
\end{equation}
For $a=20$, the ground state changes from $\left|0,0\right>$ to $\left|0,-1\right>$ at $\omega_c/\omega_0=0.2222$. Contrary, there are no level crossings for the QD (see, e.g., Fig. 1 in Ref. \onlinecite{Tanaka1}), and the level $\left|0,0\right>$ remains the ground state for all magnitudes of the magnetic field. Of course, this is immediately seen from Eqs.~\eqref{Crossing0} and \eqref{Crossing1} too where for $a\equiv0$ the transition takes place at the infinitely high intensities $B$. Limiting case of the latter equation for the large $a$ (what corresponds, as stated above, to the strictly 1D rings \cite{Buttiker1}) shows that the distance between the two consecutive crossings $\Delta\omega_m^\times\equiv\omega_{m-1}^\times-\omega_m^\times$ is exactly equal to the unit flux change $h/e$ through the effective ring area $\pi\rho_V^2$, as expected, Ref.~\onlinecite{Buttiker1}. The same is approximately true in the opposite limiting case of the small $a$ and quite large $|m|$. These crossings are an essential feature of all ring-like structures with nonsimply connected topology. To explain their emergence and physical meaning,  we point out that the states with the larger negative $m$ are located further from the origin $\rho=0$. This is immediately seen from the expression for the mean radius $\rho_{a;nm}\equiv\left<nm\right|\rho^2\left|nm\right>^{1/2}$ of the state $\left|n,m\right>$, which for the VD becomes
\begin{equation}\label{Meanradius2}
\rho_{a;nm}=\left[2\left(2n+\sqrt{m^2+a}+1\right)\right]^{1/2}\rho_{eff}.
\end{equation}
Thus, the lowest ($n=0$) circularly symmetric, $m=0$, level in the absence of the field, $\omega_c=0$, is localized mainly around the minimum of the VD and is the lowest lying state. As a result, the growing magnetic intensity pushes it closer to the centre and away from the lowest section of $U(\rho)$ increasing in this way its energy. Note that this pace of the energy change is faster for the VD as compared to its QD counterpart \cite{Tanaka1} since there is a joint influence of the magnetic intensity and the inner potential wall in the former case  while in the latter geometry, it is the field $B$ only that pushes the energy upwards. In turn, electrons in the states with the negative $m$ at the small and moderate $B$ are forced to move to the zero of the electrostatic potential with the corresponding decrease of their energies. Similar lowering is characteristic for the QD too; \cite{Tanaka1} however, at $\omega_c=0$, the energy difference between the corresponding levels is smaller for the VD, as it directly follows from Eq.~\eqref{Energy0}. Thus, for the ring structure, the combined influence of its repulsive center and the magnetic field is able to close this gap at the finite intensity $B_m^\times$, which is a function of the quantum number $m$ and the antidot strength $a$, while no level crossings are observed for the simply connected configuration of the QD \cite{Tanaka1} since the ``lone" magnetic field is not strong enough to force the levels to ``meet" each other. Optical detection of these crossings was reported in state-of-the-art FIR experiment. \cite{Lorke1} Here, we provide a detailed analysis of the interaction of the quantum ring with the optical field, which will be described by the total absorption coefficient $\alpha(\omega)$ with its linear $\alpha^{(1)}(\omega)$ and cubic $\alpha^{(3)}(\omega)$ components \cite{Boyd1}
\begin{equation}\label{TotalAbsoprtion0}
\alpha(\omega)=\alpha^{(1)}(\omega)+\alpha^{(3)}(\omega),
\end{equation}
where
\begin{subequations}\label{TotalAbsoprtion1}
\begin{eqnarray}\label{LinearAbsoprtion1}
\alpha^{(1)}(\omega)&=&\omega\sqrt{\frac{\mu}{\varepsilon_R}}\frac{NT_{if}^2}{\hbar}\frac{\Gamma_{if}}{\left(\omega_{if}-\omega\right)^2+\Gamma_{if}^2}\\
\label{NonlinearAbsoprtion1}
\alpha^{(3)}(\omega)&=&-\omega\sqrt{\frac{\mu}{\varepsilon_R}}\frac{2I}{n_{\rm r}\varepsilon_0c}\frac{NT_{if}^4}{\hbar^3}\frac{\Gamma_{if}}{\left[\left(\omega_{if}-\omega\right)^2+\Gamma_{if}^2\right]^2}.
\end{eqnarray}
\end{subequations}
Here, $\varepsilon_0= 8.85...\times10^{-12}$ F/m is the permittivity of free space, $c$ is the speed of light, $\mu$ is the permeability of the system, $n_{\rm r}$ is the medium refractive index, $\varepsilon_R=n_{\rm r}^2\varepsilon_0$ is the real part of the permittivity, $N$ is a carrier density, $I$ is the optical power per unit area, $\omega_{if}=(E_f-E_i)/\hbar$, and $\Gamma_{ij}$ is a relaxation rate between the initial ($i$) and final ($f$) states with their corresponding energies $E_i$ and $E_f$, which, in our case, are given by Eq.~\eqref{Energy0}, while 
\begin{equation}\label{DipoleElement1}
T_{if}=e\left<i|\rho e^{\pm i\varphi}|f\right>
\end{equation}
is a dipole transition matrix element between the corresponding levels. Note that the cubic contribution is proportional to the intensity of the optical radiation $I$ and is always negative decreasing in this way the total absorption. Equations~\eqref{TotalAbsoprtion0} -- \eqref{DipoleElement1} were derived under the assumption that the wavelength of the incident light is much larger than the dimensions of the system (electric dipole approximation), and different signs in Eq.~\eqref{DipoleElement1} correspond to the left or right circular polarization with the electric vector of the optical field being perpendicular to the plane of the ring (TM polarization).

Plugging the expressions for the initial $\left|i\right>\equiv\left|nm\right>$ and final  $\left|f\right>\equiv\left|n'm'\right>$ states from Eq.~\eqref{Wavefunction0} into $T_{ij}$ and carrying out a polar integration, one immediately obtains the selection rule for the magnetic quantum number $m$
\begin{equation}\label{SelectionRule2}
\Delta m\equiv m'-m=\pm1,
\end{equation}
which is the same as for the QD. \cite{Liu1,Geerinckx1} Radial integral
\begin{eqnarray}
&&\int_0^\infty\!\rho^2R_{nm}\left(a;\rho\right)R_{n'm'}\left(a;\rho\right)d\rho\nonumber\\
&&=2^{1/2}\rho_{e\!f\!f}\left[\frac{n!}{\Gamma\!\left(n+\sqrt{m^2+a}+1\right)}\frac{n'!}{\Gamma\!\left(n'+\sqrt{{m'}^2+a}+1\right)}\right]^{1/2}\nonumber\\
\label{integral1}
&&\times\int_0^\infty e^{-x}x^{(\sqrt{m^2+a}+\sqrt{{m'}^2+a}+1)/2}L_n^{\sqrt{m^2+a}}\left(x\right)L_{n'}^{\sqrt{m'^2+a}}\left(x\right)dx
\end{eqnarray}
in the expression for the dipole matrix element $T_{nn'}^{mm'}$ (where $m'=m\pm1$) in the case of the QD, $a=0$, produces the well known selection rules \cite{Liu1,Geerinckx1} for the transitions between different Landau subbands $n$ and $n'$; namely, 
\begin{equation}\label{SelectionRule1}
|n'-n|=0,1,\quad a=0.
\end{equation}
The same requirement applies for the free, $\omega_0=a=0$, particle in the uniform magnetic field. \cite{Dingle1} However, for the quantum ring, $a\omega_0\ne0$, the rigorous demand of Eq.~\eqref{SelectionRule1} is lifted as an explicit evaluation reveals \cite{Prudnikov1}
\begin{eqnarray}
%&&\int_0^\infty e^{-x}x^{(\lambda+\lambda'+1)/2}L_n^{\lambda}\left(x\right)L_{n'}^{\lambda'}\left(x\right)dx\nonumber\\
%&&=\frac{(1+\lambda)_n\left(-\frac{\lambda-\lambda'+1}{2}\right)_{n'}\Gamma\left(1+\frac{\lambda+\lambda'+1}{2}\right)}{n!n'!}\nonumber\\
\int_0^\infty \!\!e^{-x}x^{(\lambda+\lambda'+1)/2}L_n^{\lambda}\left(x\right)\!L_{n'}^{\lambda'}\left(x\right)dx=\frac{(1+\lambda)_n\!\left(-\frac{\lambda-\lambda'+1}{2}\right)_{n'}\Gamma\!\!\left(1\!+\frac{\lambda+\lambda'+1}{2}\right)}{n!n'!}\nonumber&&\\
\label{integralValue1}
\times{}_3F_2\!\!\left(\!-n,1\!+\frac{\lambda\!+\!\lambda'\!+\!1}{2},1\!+\frac{\lambda\!-\!\lambda'\!+\!1}{2};\!\lambda\!+\!1,1\!+\!\frac{\lambda\!-\!\lambda'\!+\!1}{2}-\!n';1\!\right),\, n\le n'&&
\end{eqnarray}
(for $n>n'$ the primed and unprimed indices $n$ and $\lambda$ in the right-hand side of Eq. (\ref{integralValue1}) should be interchanged), where $(\alpha)_n$ is a Pochhammer symbol and the generalized hypergeometric function \cite{Bateman1} $_3F_2(\alpha_1,\alpha_2,\alpha_3;\beta_1,\beta_2;x)$ for the negative integer $\alpha_i$ ($i=1,2$ or $3$) reduces to the polynomial of the order $-\alpha_i$. Accordingly, the transitions between different Landau subbands without any restriction on their radial numbers are possible; \cite{Bogachek1} in particular, for the transitions involving the lowest Landau state, $n=0$, the dipole matrix element is
\begin{equation}\label{TransitionElement1}
T_{0n'}^{m,m\pm1}=2^{1/2}e\rho_{e\!f\!f}\frac{\Gamma\!\!\left(1+\frac{\lambda+\lambda'+1}{2}\right)}{\left[n'!\,\Gamma(\lambda+1)\,\Gamma(n'+\lambda'+1)\right]^{1/2}}\,\frac{\Gamma\!\!\left(n'-\frac{\lambda-\lambda'+1}{2}\right)}{\Gamma\!\!\left(-\frac{\lambda-\lambda'+1}{2}\right)}
\end{equation}
with $\lambda=(m^2+a)^{1/2}$ and $\lambda'=[(m\pm1)^2+a]^{1/2}$. Below, we will be interested in the intraband transitions between the lowest crossing levels, $n=n'=0$, what simplifies the square of the transition matrix element in Eqs.~\eqref{TotalAbsoprtion1} to
\begin{equation}\label{TransitionElement2}
(T_{00}^{m,m\pm1})^2=2e^2\rho_{e\!f\!f}^2\frac{\Gamma^2\!\!\left(1+\frac{\lambda+\lambda'+1}{2}\right)}{\Gamma(\lambda+1)\,\Gamma(\lambda'+1)}.
\end{equation}
\section{Results and discussion}\label{sec_3}
\begin{figure}
\centering
\includegraphics[width=0.99\columnwidth]{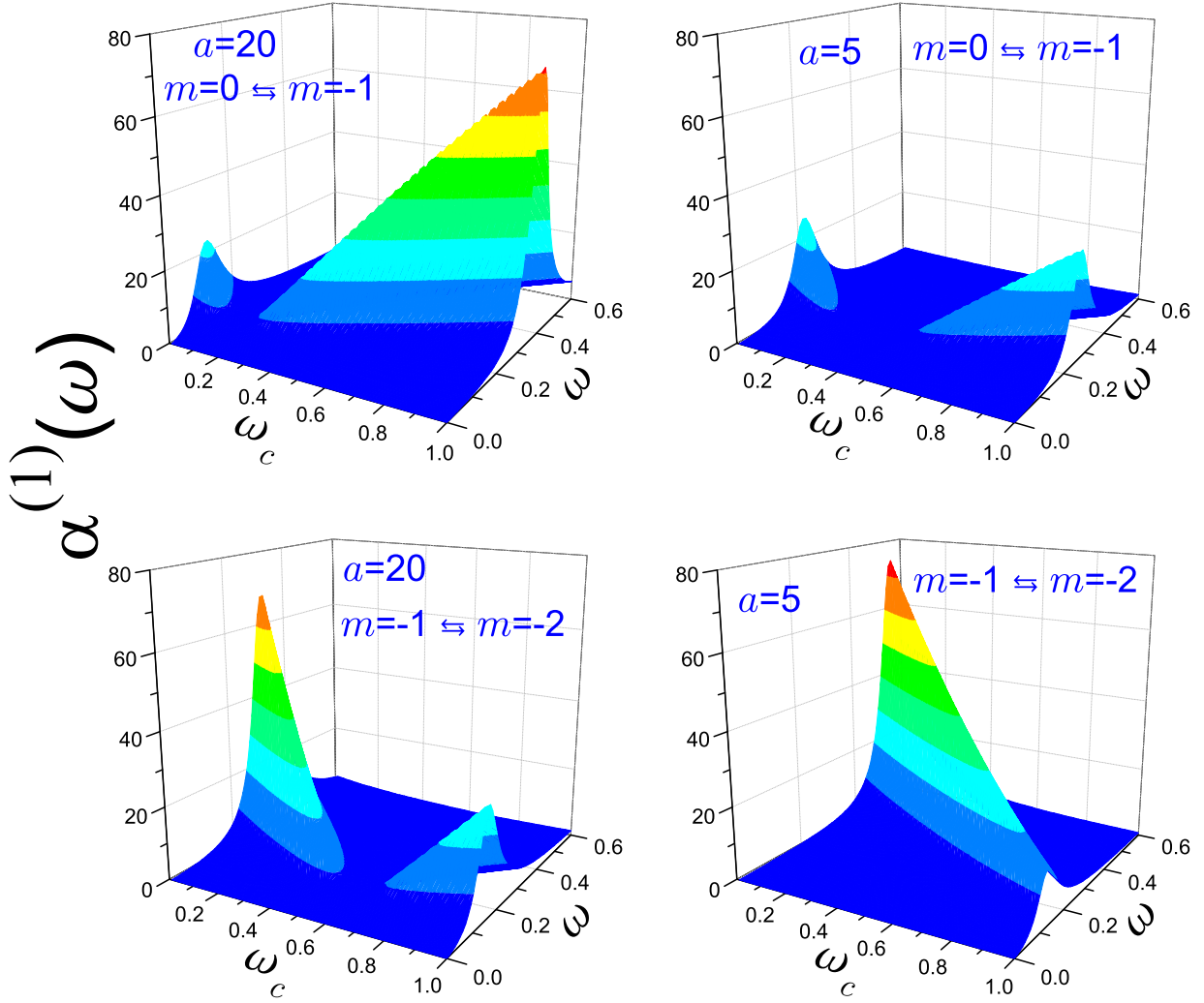}
\caption{\label{Fig3}
Linear optical absorption $\alpha^{(1)}(\omega)$ expressed in units of $\sqrt{\frac{\mu}{\varepsilon_R}}N\frac{e^2}{\hbar}\rho_0^2$ as a function of the light $\omega$ and cyclotron $\omega_c$ frequencies (both expressed in units of $\omega_0$) for the antidot strength $a=20$ (left panels) and $a=5$ (right panels). Two upper figures show dependencies for the transition between levels $\left|0,0\right>$ and $\left|0,-1\right>$ while the two lower plots are for the states $\left|0,-1\right>$ and $\left|0,-2\right>$. The relaxation rate $\Gamma_{ij}$ is assumed to be $\Gamma_{ij}=\frac{1}{20}$ (in units of $\omega_0$). In the lower right panel a minimum of the absorption peak is located at $\omega_c>1$ and is not shown in the figure.}
\end{figure}

First, we analyze the linear absorption.  Optical spectrum will consist of the infinitely many lines governed by the selection rule, Eq.~\eqref{SelectionRule2}, with their intensities described by Eq.~\eqref{LinearAbsoprtion1}. Fig.~\ref{Fig3} depicts coefficient $\alpha^{(1)}$ as a function of the frequencies $\omega$ and $\omega_c$ for the transitions between the states $\left|0,0\right>$ and $\left|0,-1\right>$ (two top panels) and $\left|0,-1\right>$ and $\left|0,-2\right>$ (lower panels) for the two different strengths of the antidot $a$, which, according to Eqs.~\eqref{Radius1} and \eqref{InnerOuterRadii}, determines rings of the different radii. It is seen that tuning the magnetic field has a dramatic effect on the optical properties with the absorption being severely suppressed at the intensities $B$ corresponding to the level crossing. Physically, this enlightenment of the sample is explained by the fact that no electronic transitions are possible at the crossing between the two degenerate levels; accordingly, no absorption takes place and the incident light passes unattenuated through the surface of the 2D ring. Small nonzero coefficient $\alpha^{(1)}$ at $\omega_c\sim\omega_m^\times$ observed in Fig.~\ref{Fig3} is explained by the levels' broadening with its magnitude being described by the relaxation rate $\Gamma_{ij}$. The broadening can be affected by different factors; in particular, the growing temperature increases the coefficient $\Gamma_{ij}$. As a representative example, in our calculations we kept the value of $\Gamma_{ij}$ equal to $1/20$ of $\omega_0$, which, as mentioned above, is the most convenient unit of measuring the frequency. It follows from Eq.~\eqref{LinearAbsoprtion1} that smaller (larger) relaxation rates lead to the increase (decrease) of the peak value and to the narrowing (broadening) of the corresponding resonance. As the magnetic field changes from $\omega_m^\times$ in either direction,  the absorption increases since the electron by absorbing the photon can make a transition between the states whose energies move away from each other. For the same reason, the optical frequency $\omega$, at which the maximum absorption occurs, grows together with $|\omega_c-\omega_m^\times|$.

To support mathematically physical explanation of the results exhibited by Fig.~\ref{Fig3}, we present the linear absorption as a function of two variables
\begin{equation}\label{AbsorptionExample1}
\alpha^{(1)}(\omega,\omega_c)=f(\omega_c)\frac{\omega}{[\omega-\omega_{if}(\omega_c)]^2+\Gamma_{if}^2},
\end{equation}
where the nonnegative $\omega_{if}(\omega_c)$ reaches its zero minimum at $\omega_m^\times$ while the slowly varying function $f(\omega_c)$ 
\begin{equation}\label{Function_f}
%f(\omega_c)=2\Gamma_{if}\frac{C}{[\omega_0^2+\omega_c^2/4]^{1/2}}
f(\omega_c)=2\Gamma_{if}\frac{C}{\left(\omega_0^2+\omega_c^2/4\right)^{1/2}}
\end{equation}
with a positive constant $C>0$ accommodating all other coefficients from the right-hand side of Eq.~\eqref{LinearAbsoprtion1}, takes at the same frequency a finite positive value. As a function of $\omega$, the absorption reaches its maximum of 
\begin{equation}\label{AbsorptionMax1}
\alpha_{max}^{(1)}(\omega_c)=f(\omega_c)\frac{\left[\omega_{if}^2(\omega_c)+\Gamma_{if}^2\right]^{1/2}}{\left(\left[\omega_{if}^2(\omega_c)+\Gamma_{if}^2\right]^{1/2}-\omega_{if}(\omega_c)\right)^2+\Gamma_{if}^2}
\end{equation}
at $\omega_{max}=\left[\omega_{if}^2(\omega_c)+\Gamma_{if}^2\right]^{1/2}$. This last dependence at $\omega_{if}=0$ (i.e., at $\omega_c=\omega_m^\times$), when $\omega_{max}=\Gamma_{if}$, has a minimum of
\begin{equation}\label{AbsorptionMaxMin1}
\alpha_{max}^{(1)}\!\left(\omega_m^\times\right)=\frac{C}{\left[\omega_0^2+\left(\omega_m^\times/2\right)^2\right]^{1/2}}.
\end{equation}
On the increase of the magnetic field from $\omega_m^\times$, the extremum $\alpha_{max}^{(1)}(\omega_c)$ grows until the presence of the envelope potential $f(\omega_c)$ causes it to saturate to
\begin{equation}\label{LinearLimit1}
%\alpha_{max}^{(1)}\!\left(\omega_c\rightarrow\infty\right)=2C\frac{\lambda-\lambda'+1}{\Gamma_{if}}.
\alpha_{max}^{(1)}\!\left(\omega_c\right)\xrightarrow[\omega_c\rightarrow\infty]{}2\,C\frac{\lambda-\lambda'+1}{\Gamma_{if}}.
\end{equation}
However, this flattening takes place at the high intensities $B$, so, in our range of interest $\omega_m^\times\le\omega_c\lesssim (2\div3)\,\omega_0$, one can safely  assume that the maximum of the absorption is a linearly increasing function of the magnetic field.

Comparing left and right panels in Fig.~\ref{Fig3}, one sees the dependence of the critical magnetic field $B_m^\times$ on the dimensions of the ring; namely, for the smaller rings [right panels, cf. Eq.~\eqref{Radius1}], the larger $\omega_c$ are needed to reach the minimum of the absorption since the magnetic radius $\rho_B=\left[\hbar/(eB)\right]^{1/2}$ should be commensurate there with $\rho_V$. For the parameters from the lower right panel $\omega_{-1}^\times=1.3189\,\omega_0$, and the corresponding global minimum is not shown.

\begin{figure}
\centering
\includegraphics[width=0.95\columnwidth]{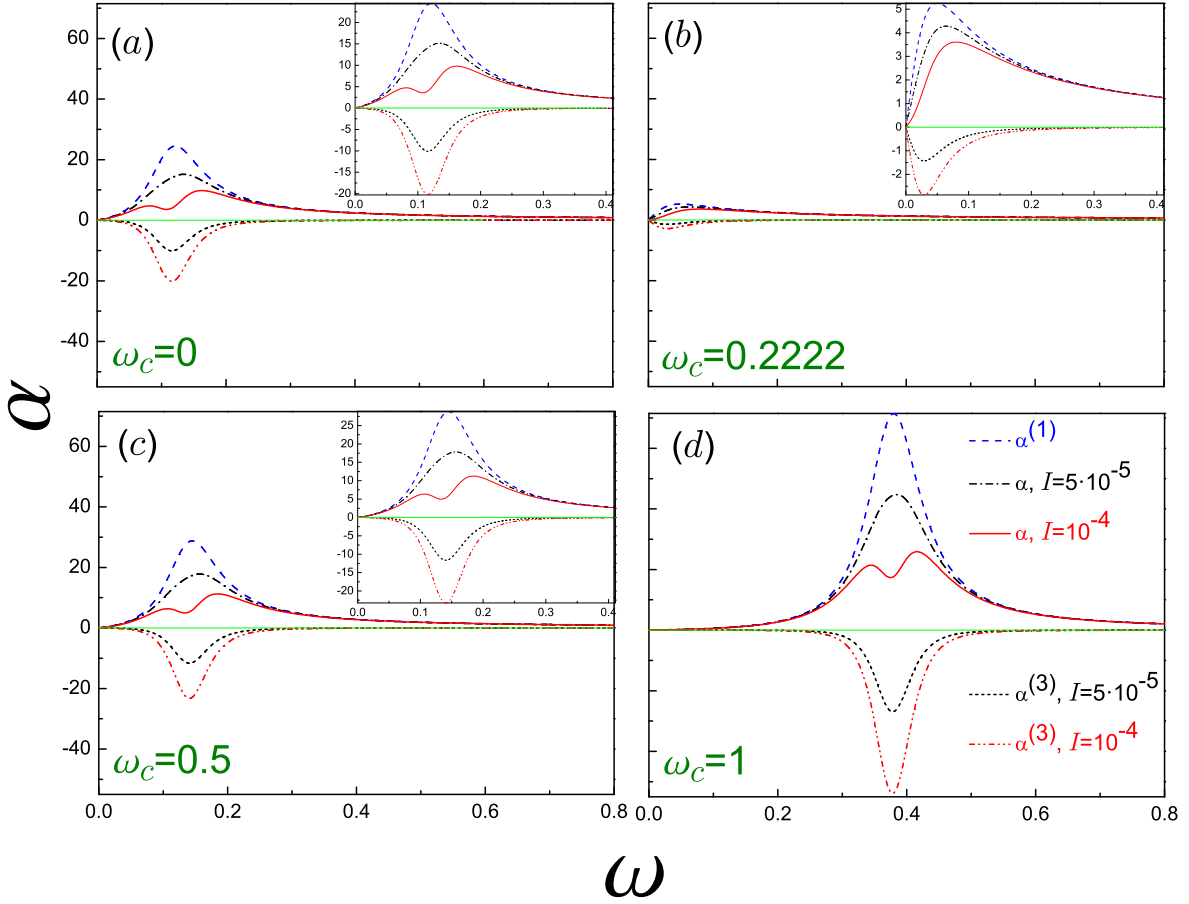}
\caption{\label{Fig4}
Linear $\alpha^{(1)}$, cubic $\alpha^{(3)}$ and total $\alpha=\alpha^{(1)}+\alpha^{(3)}$ optical absorption coefficients for the transition between the levels $\left|0,0\right>$ and $\left|0,-1\right>$ as a function of the frequency of the optical field $\omega$ for the Volcano ring with $a=20$, $\Gamma=\frac{1}{20}$ and several magnetic fields where panel (a) is for $\omega_c=0$, (b) $\omega_c=0.2222$, (c) $\omega_c=0.5$ and (d) $\omega_c=1$. Dashed lines depict linear coefficient $\alpha^{(1)}$, dotted (dash-dot-dotted) curves are for the cubic absorption coefficient $\alpha^{(3)}$ for the intensity $I=5\times10^{-5}$ ($I=10^{-4}$), and dash-dotted (solid) lines show the total coefficient $\alpha$ for the same intensity. Frequencies are measured in units of $\omega_0$, absorption coefficients  - in units of $\sqrt{\frac{\mu}{\varepsilon_R}}N\frac{e^2}{\hbar}\rho_0^2$, and intensities $I$ - in units of $n_{\bf r}\varepsilon c(\hbar\omega_0)^2/(e\rho_0)^2$. For clarity, insets in panels (a)-(c) show the same dependencies in the larger format. Note different $\alpha$ scales in each inset. }
\end{figure}

Having learned the features of the linear absorption, let us turn next to its cubic counterpart and influence of its interaction with the linear term on the full spectrum. Fig.~\ref{Fig4} shows linear $\alpha^{(1)}(\omega)$, cubic $\alpha^{(3)}(\omega)$ and total $\alpha(\omega)$ absorptions for several magnetic fields $\omega_c$. As the cubic contribution is negative, it decreases the total coefficient. At quite large intensities, another additional maximum on the $\alpha-\omega$ dependence emerges separated by the minimum  from the peak that existed in the linear case (which in the following we will call a ``linear maximum"). Existence of these extrema has been known for a while;\cite{Ahn1} however, despite a lot of research,\cite{Wang1,Liu3,Baghramyan1,Guo1,Ozturk1,Kavruk1} no consistent theory of this phenomenon is known to the authors. Here, we close this gap by analysing the critical intensity $I_{cr}$, at which these resonances emerge, its location $\omega_{cr}$ on the $\omega$ axis and their dependence on the magnetic field $\omega_c$. To make the results as generic as possible, we introduce the function
\begin{equation}\label{FunctionG1}
g(x)=\frac{x}{(x-x_0)^2+1}-A\,\frac{x}{\left[(x-x_0)^2+1\right]^2}
\end{equation}
with the nonnegative coefficients $x_0$ and $A$: $x_0\ge0$, $A\ge0$. Comparing it with the expression for the total absorption, Eq.~\eqref{TotalAbsoprtion0}, one sees that the function $g(x)$ represents dimensionless coefficient $\alpha$ if  the variable $x$ in Eq.~\eqref{FunctionG1} substitutes the frequency $\omega$ measured in units of the level broadening $\Gamma_{if}$, $x=\omega/\Gamma_{if}$. In the same way, the factor $x_0$ stands for the dimensionless energy difference $\omega_{if}$ while the coefficient $A$ is essentially an ``effective" optical intensity since $A\sim T_{if}^2I/\Gamma_{if}^2$. All other coefficients that appear in Eqs.~\eqref{LinearAbsoprtion1} and \eqref{NonlinearAbsoprtion1} are insignificant for our present analysis and they have been absorbed by the proportionality coefficient between $\alpha(\omega)$ and $g(x)$. In this way, the results presented below are equally applicable to other physical systems.\cite{Ahn1,Wang1,Liu3,Baghramyan1,Guo1,Ozturk1,Kavruk1} Thus, the emergence, evolution and number of the extrema in the absorption $\alpha(\omega)$ are determined by their counterparts in the function $g(x)$. Accordingly, in our description below we will interchangeably use, if it does not cause any confusion, the functions $g(x)$ and $\alpha(\omega)$ as well as the frequencies $\omega$ and $\omega_{ij}$, on the one hand, and the  variables $x$ and $x_0$, on the other one. In the linear case, $A=0$, the only extremum, which is a maximum of the magnitude 
\begin{equation}\label{Gmax1}
g_{max}^{(1)}\rvert_{A=0}\equiv g(x_{max})\rvert_{A=0}=\frac{(x_0^2+1)^{1/2}}{[(x_0^2+1)^{1/2}-x_0]^2+1},
\end{equation}
is located at $x_{max}\rvert_{A=0}=(x_0^2+1)^{1/2}$. We are interested in finding a critical optical intensity $A_{cr}$ at which the new extrema that were absent in the linear regime emerge on the $g-x$ (or, correspondingly, $\alpha-\omega$) characteristics, and their locations $x_{cr}$ (or $\omega_{cr}$). To do so, one needs to zero a derivative of the total absorption $g(x)$ from Eq.~\eqref{FunctionG1} with respect to the normalized frequency $x$. Solutions of the resulting quartic equation with the intensity- and $x_0$-dependent coefficients define locations and magnitudes of the corresponding dips and peaks. In general, these {\it analytical} results are very unwieldy and hard to comprehend and get  an easy useful information.  Moreover, for finding $x_{cr}$, one needs to zero a second derivative of the function $g(x)$ what complicates even more the analytical discussion. However, situation simplifies drastically for some limiting cases. For example, for $x_0=0$, one immediately finds that 
\begin{equation}\label{CriticalIntensity1}
%x_0=0\Rightarrow\left\{\begin{array}{ccc}
%A_{cr}&=&1\\
%x_{cr}&=&0.
%\end{array}
%\right.
\left.\begin{array}{ccc}
A_{cr}&=&1\\
x_{cr}&=&0
\end{array}
\right\}\quad{\rm at}\quad x_0=0.
\end{equation}
Corresponding ``linear" maximum with its height of
\begin{equation}\label{Gmax2}
g_{max}^{(1)}\rvert_{A=1,\,x_0=0}=\frac{3\cdot3^{1/2}}{16}
\end{equation}
is located at $x_{max}^{(1)}\rvert_{A=1,\,x_0=0}=3^{1/2}$. This allows to introduce, along with the critical parameters $A_{cr}$ and $x_{cr}$, another quantity that is a frequency difference between $x_{max}^{(1)}$ and the corresponding $x_{cr}$
\begin{equation}\label{DeltaCritical1}
\Delta_{cr}(x_0)=x_{max}^{(1)}-x_{cr}.
\end{equation}
Note that the case $x_0=0$ is unique in a sense that the growing optical power $A$ gives birth to only {\it one} intensity-induced extremum; namely, for any $A>1$, the total optical absorption becomes negative with the frequency increasing from zero. This plunge is explained by the stronger magnitude of the cubic contribution for the small and moderate  $\omega$, cf. Eqs.~\eqref{TotalAbsoprtion1} at $\omega_{if}=0$. However, the nonlinear absorption is much sharper localized on the $\omega$ axis. Mathematically, this physical effect is expressed by the additional power of $\left(\omega_{if}-\omega\right)^2+\Gamma_{if}^2$ in the denominator of Eq.~\eqref{NonlinearAbsoprtion1} as compared to the one from Eq.~\eqref{LinearAbsoprtion1}. Accordingly, after reaching the negative minimum at some frequency, the total coefficient $\alpha$ begins to grow as the influence of the linear term starts to dominate. At still higher frequencies, this domination becomes total.
\begin{figure}
\centering
\includegraphics[width=0.95\columnwidth]{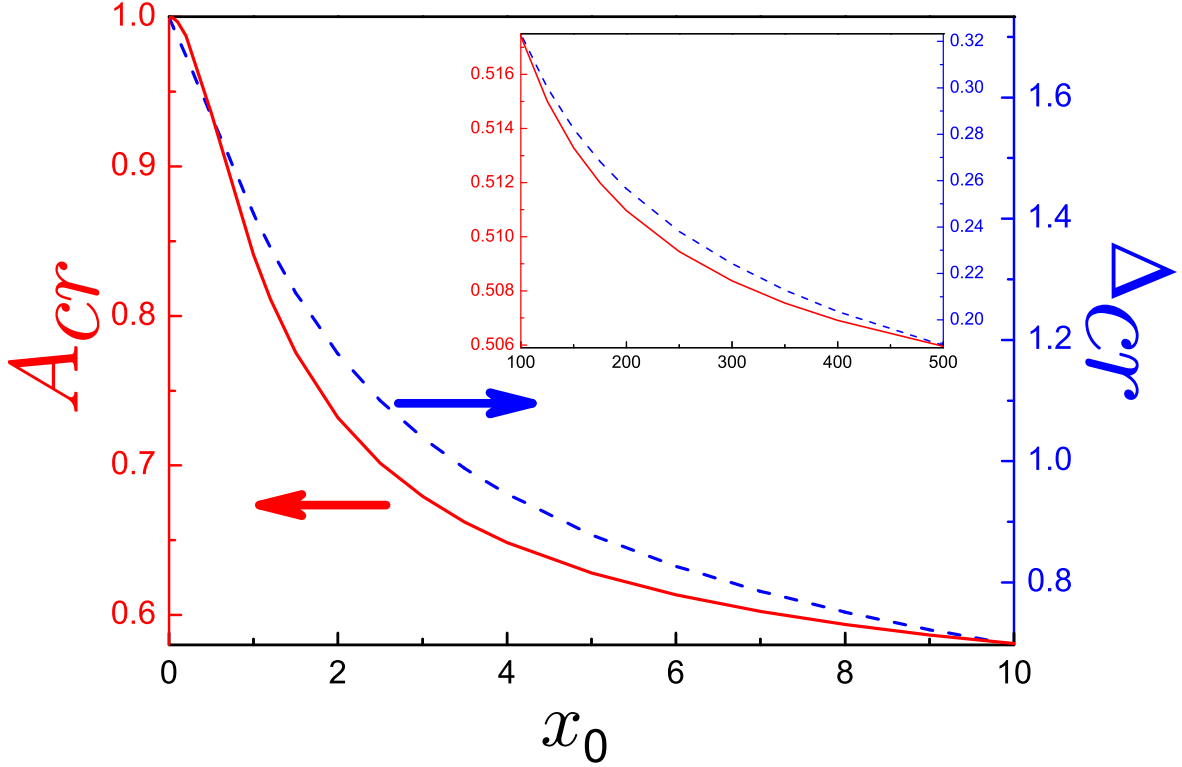}
\caption{\label{Fig5}
Critical intensity $A_{cr}$ (solid line, left axis) and difference $\Delta_{cr}$ from Eq.~\eqref{DeltaCritical1} (dashed curve, right axis) as a function of the normalized energy difference $x_0$. The inset shows the same quantities for the large $x_0$. Note different axes ranges in the main figure and the inset.}
\end{figure}

For any nonzero $x_0$, the large optical power $I$ produces two additional extrema, as it is seen from panels (a), (c) and (d) of Fig.~\ref{Fig4}. If it is quite small, $x_0\ll1$, one can get simple analytical results; namely, in this case the critical frequency is small too, $x_{cr}\ll1$; and, accordingly, in its neighborhood, the function $g(x)$ from Eq.~\eqref{FunctionG1} is represented as
\begin{equation}\label{FunctionG2}
g(x)=x\!\left[1-(x-x_0)^2\right]-A\,x\!\left[1-2(x-x_0)^2\right],\qquad x,x_0\ll1.
\end{equation}
Zero of its derivative, $g'(x)=0$, results in a quadratic equation whose vanishing determinant immediately produces the desired quantities
\begin{equation}\label{CriticalIntensity2}
%x_0\ll1\Rightarrow\left\{\begin{array}{ccc}
%A_{cr}&=&1-\frac{1}{3}x_0^2\\
%x_{cr}&=&\frac{2}{3}x_0\\
%g_{cr}&=&\frac{8}{27}x_0^3,
%\end{array}
%\right.
\left.\begin{array}{ccc}
A_{cr}&=&1-\frac{1}{3}\,x_0^2\\
x_{cr}&=&\frac{2}{3}\,x_0\\
g_{cr}&=&\frac{8}{27}\,x_0^3
\end{array}
\right\}\quad{\rm at}\quad x_0\ll1,
\end{equation}
where $g_{cr}\equiv g(x_{cr})$. In the opposite limit of the very large energy differences, $x_0\gg1$, three corresponding extrema are located very close to each other. Accordingly, the quartic equation can be separated into two parts: the first one groups the large terms, and its simple solution produces triply degenerate resonance at $x=x_0$. The second part unites the small factors and can be considered as a perturbation to the first one. This disturbance lifts just mentioned degeneracy. After some algebra that retains the relevant powers of $x_0$, one gets
\begin{equation}\label{CriticalIntensity3}
%x_0\gg1\Rightarrow\left\{\begin{array}{ccc}
%A_{cr}&=&\frac{1}{2}+\frac{3}{4}\frac{1}{x_0^2}\\
%x_{cr}&=&x_0-\frac{1}{3}\frac{1}{x_0}\\
%g_{cr}&=&\frac{8}{27}x_0^3,
%\end{array}
%\right.
\left.\begin{array}{ccc}
A_{cr}&=&\frac{1}{2}+\frac{3}{8}\frac{1}{x_0^{2/3}}\\
x_{cr}&=&x_0-\frac{1}{2}\frac{1}{x_0^{1/3}}+\frac{1}{4}\frac{1}{x_0}\\
g_{cr}&=&\frac{1}{2}x_0-\frac{3}{8}x_0^{1/3}-\frac{3}{32}\frac{1}{x_0^{1/3}}
\end{array}
\right\}\quad{\rm at}\quad x_0\gg1.
\end{equation}
Since in the same limit $x_0\gg1$, the location of the linear maximum is given as $x_{max}^{(1)}\rvert_{A=A_{cr}}=x_0+\frac{1}{x_0^{1/3}}+\frac{1}{4}\frac{1}{x_0}$, we find
\begin{equation}\label{DeltaCritical2}
\Delta_{cr}=\frac{3}{2}\frac{1}{x_0^{1/3}},\quad x_0\gg1.
\end{equation}

Equations~\eqref{CriticalIntensity2} and \eqref{CriticalIntensity3} manifest that the critical optical power $A_{cr}(x_0)$, which switches additional resonances in the absorption spectrum, on the whole frequency axis changes in the limited range
\begin{equation}\label{Range1}
%\frac{1}{2}\le A_{cr}(x_0)\le1,\quad\infty\ge x_0\ge0;
\frac{1}{2}\le A_{cr}(x_0)\le1\quad{\rm for}\quad\infty\ge x_0\ge0;
\end{equation}
in particular, for the very small detuning, it from unity decreases quadratically with $x_0$; while for the very large energy gap between the initial and final states it is equal to the one half with small positive admixture that is inversely proportional to $x_0^{2/3}$. Interaction of these two asymptotics in the intermediate regime $x_0\sim1$ produces the dependence shown by the solid line in Fig.~\ref{Fig5}. It is seen that the critical intensity $I_{cr}$ monotonically decreases with the growing $\omega_{ij}$ with its speed of change being dependent on the latter. For all optical powers lying below the solid line, the total absorption exhibits only one extremum; while  for the larger intensities, i.e., for the points above the solid line in Fig.~\ref{Fig5}, the two maxima separated by the minimum are observed. The difference between the location of the linear maximum $x_{max}^{(1)}$ and the critical point $x_{cr}$ is shown by the dashed line in Fig.~\ref{Fig5}. The magnitude of $\Delta_{cr}$ smoothly descents to zero from its $3^{1/2}$ value at $x_0=0$. The inset shows the asymptotic behavior of these two values at the large $x_0$. Due to the small inverse powers of $x_0$ in Eqs.~\eqref{CriticalIntensity3} and \eqref{DeltaCritical2}, the decay of $A_{cr}$ and $\Delta_{cr}$ with the energy gap growing is quite slow.  As a final remark of this part of our discussion, let us note that at $A=1$, the intensity-induced dip is located at $x=x_0$ and its magnitude is equal to zero, as it elementary follows from Eq.~\eqref{FunctionG1}.

Applying the results developed in the above paragraphs to Fig.~\ref{Fig4}, we turn first to panel (b) where the choice of $\omega_c$ corresponds to $x_0=0$. Critical power is the largest in this case and the two intensities used by us in Fig.~\ref{Fig4} are smaller than $I_{cr}$; accordingly, for either of them only one maximum is observed in the total absorption dependence on the frequency $\omega$. Changing the magnetic field from $\omega_m^\times$ leads to the nonzero energy gap and to the concomitant decrease of the critical intensity. Consequently, the conditions for the emergence of the intensity-induced extrema are eased and, indeed, for the larger optical power from Fig.~\ref{Fig4}, they are clearly observed in each of the remaining panels while the smaller $I$ is not strong enough to produce extra dip and peak. Magnitude of the cubic contribution is the largest in the additional minimum $\omega_{min}$ and so, the detuning of the optical frequency $\omega$ from $\omega_{min}$ leads to the increase  of the total absorption. Note that the linear term decreases too; however, its pace of change with $\omega$ is, due to the additional power in the denominator of Eq.~\eqref{NonlinearAbsoprtion1}, much slower than that of the cubic part. As a result, the total absorption reaches maximum after which it gets smaller with its magnitude for the relatively large $|\omega-\omega_{min}|$ being determined mainly by the linear contribution. Note that at the large energy gap between the initial and final states the optical resonances are located very close to each other on the $\omega$ axis; accordingly, additional care should be taken in order to resolve them in the experiment.

\section{Concluding Remarks}\label{sec_4}
Analysis of the optical absorption of the quantum ring revealed that the magnitude of its linear term drastically decreases when the increasing magnetic field $\bf B$ forces the energies of the two adjacent quantum states to cross. The advantage of the chosen model of the VD allowed to derive simple analytical results and get their clear physical explanation; for example, an expression for the field $B_m^\times$ at which the levels cross describes both ``thick" and ``thin" loops. The mathematics employed permits to show the physical reasons for the sample enlightenment in the vicinity of the crossing. The interplay between linear and intensity-dependent cubic absorptions is analyzed too and the critical optical power that switches additional resonances in the spectrum is calculated; in particular, in the limiting cases of the large and small frequencies its simple analytical expressions are derived.

Effects of the electron-electron interaction are neglected in our calculations. Generalization of the Kohn theorem \cite{Kohn1} stating that the optical response of the 2D gas is independent of the Coulomb force is applicable for the the parabolic QD only and does not work for other geometries (like ours) where the mixing of the centre-of-mass and relative motions may produce additional resonances in the FIR spectrum. Nevertheless, experiment showed \cite{Lorke1} that for the displaced parabola potential from Eq.~\eqref{DisplacedParabola1}, the single-particle states are a quite accurate basis for the description of the many-particle states and excitations and that the measured FIR resonance positions for one electron are very similar to those for the two like charges in the ring. On the basis of the calculation of the linear FIR  spectra  for two electrons in the potential from Eq.~\eqref{DisplacedParabola1}, it was argued \cite{Hu1} that inclusion of the Coulomb scattering should produce additional crossings, which, however, can be resolved only in the even higher resolution experiments. Moreover, for the similar semiconductor structures, the independent electron model of the VD \cite{Tan1,Tan2,Tan3} was very successful in explaining transport experiments in GaAs/Al$_x$Ga$_{1-x}$As rings. \cite{Liu2,Mailly1} Accordingly, we believe that all the features discussed above should survive in the presence of the multi-particle interactions. Discussion of the influence of the electron-electron scattering on the optical properties of the quantum rings is far from being terminated. \cite{Emperador1,Climente1}

Above, to keep our results as general as possible we have used dimensionless units. Let us make some numerical estimates. Typical radius of the AlGaAs-GaAs quantum rings used in the experiments \cite{Liu1,Fuhrer1} is around 100 nm. Assuming this value for the mean radius $\rho_V$ in our model of the VD with $a=20$ and $m^\ast=0.067m_e$ ($m_e$ being a free electron mass), one calculates QD confining energy $\hbar\omega_0$ of $0.25$ meV. Accordingly, the drastic decrease of the absorption for the transitions between the states with $m=0$ and $m=-1$ should take place at the magnetic field $B_0^\times\sim32.7$ mT, and the one corresponding to the transitions $m=-1\leftrightarrow m=-2$  - at $B_{-1}^\times\sim98.2$ mT.  In the experiment with self-assembled InAs quantum rings, \cite{Lorke1} they had much smaller radius $\rho_V\sim18$ nm. Then, the critical magnetic field $B_0^\times\sim8$ T corresponds to the antidot strength $a\sim6$ with $\hbar\omega_0\sim4.2$ meV. Apparently, the steepness $\hbar\omega_0$ can be controlled (besides growth process) by the gate voltage applied to the structure.\cite{Liu1} Thus, application of the constant external fields to the ring allows to change its magneto-optical properties in a wide range.

%\section{Acknowledgements}\label{sec_6}
%\begin{acknowledgments}
\acknowledgments
This project was supported by Deanship of Scientific Research, College of Science Research Center, King Saud University.
%\end{acknowledgments}

\bibliography{aipsamp}

\end{document}